\documentclass[10pt]{olplainarticle}
\usepackage{textgreek,graphicx}
\usepackage{rotating}
\usepackage{float}
\usepackage{xr}
\usepackage{hyperref}
\usepackage[normalem]{ulem}
\usepackage{tabularx}
\usepackage{wrapfig}
\usepackage{multicol}
\usepackage{graphicx}
\graphicspath{ {./images/} }
\usepackage{subcaption}
\usepackage{float}
\usepackage{hyperref}

\newtheorem{definition}{\bf Definition}[section]

\usepackage{stmaryrd}
\expandafter\def\csname opt@stmaryrd.sty\endcsname
{only,shortleftarrow,shortrightarrow}
\usepackage{extpfeil}
\usepackage{ctable} 

\usepackage{moreverb} 

\title{Protein container disassembly pathways depend on geometric design}

\author[1]{Q. Roussel}
\author[2]{S. Benbedra}
\author[3,*]{R. Twarock}
\affil[1]{Télécom Paris, Institut polytechnique de Paris, Palaiseau, France}
\affil[2]{École des Ponts ParisTech, Marne-La-Vallée, France}
\affil[3]{Departments of Mathematics and Biology, University of York, York YO10 5GE, UK}

\affil[*]{corresponding author: reidun.twarock@york.ac.uk}


\begin{abstract}
The majority of viruses are organised according to the structural blueprints of the seminal Caspar-Klug theory. However, there are a number of notable exceptions to this geometric design principle. Prominent examples are the cancer-causing papilloma viridae and the \textit{de novo} designed AaLS cages that exhibit non-quasiequivalent capsid structures with protein numbers excluded by Caspar-Klug theory. The biophysical properties of these geometrically distinct architectures and the fitness advantages driving their evolution are currently unclear. We investigate here the resilience to fragmentation and disassembly behaviour of these capsid geometries by introducing a percolation theory on weighted graphs. We show that these cage architectures follow one of two distinct disassembly pathways, preferring either hole formation or capsid fragmentation. This suggests that preference for specific disassembly scenarios could be a driving force for the evolution of the non Caspar-Klug protein container architectures. 
\end{abstract}

\begin{document}

\maketitle



















\section{Introduction}

The majority of viruses package their genomes into icosahedral protein containers, called viral capsids, that provide protection for their genetic material during rounds of infection. These containers must be stable enough to protect their genetic cargo, yet also sufficiently unstable to enable its timely release at the appropriate time in the viral life cycle. Recently, we showed that capsids organised according to distinct types of surface lattices can have widely different resilience to fragmentation \cite{brunk2021percolation}. This analysis was limited to capsids abiding by the quasiequivalence principle introduced by Caspar and Klug \cite{caspar1962physical}, i.e. to those in which protein subunits make the same type of interaction across the entire capsid surface. A comparative analysis of three quasiequivalent surface lattice architectures -- a triangulation, a rhomb and a kite tiling -- was carried out, revealing different propensities to fragment for these distinct surface lattice types. 

The majority of icosahedral viruses are quasiequivalent, including those following Archimedean surface lattice architectures \cite{twarock2019structural}, and they can therefore all be studied with the approach reported earlier \cite{brunk2021percolation}. However, it is not directly applicable to the non-quasiequivalent architectures, in which protein units make several distinct types of interactions with other capsid proteins. A prominent example are the cancer-causing papilloma viridae, which exhibit two distinct types of interaction mediated by the C-terminal arms of the protein units. 

We address here the question whether such non-quasiequivalent cage architectures have stability properties, in terms of their propensity to fragment and their disassembly pathways, that differ from those of the quasiequivalent cage structures. For this, we generalise the percolation theory for quasiequivalent surface structures in Ref. \cite{brunk2021percolation} in two ways. First, we introduce a percolation theory approach based on weighted graphs, which tracks the fragmentation threshold in dependence of the "energy" equivalent of the total number of bonds removed, rather than the number of bonds removed as had previously been the case. Second, we adapt our computational strategy to correct the "energy" of protein units in disassembly intermediates to account for partially broken bonds. Both is required to adequately model the non-quasiequivalent surface architecture of these viruses because distinct interaction types make different contribtions to container disassembly. 

We start by introducing our mathematical model of papillomavirus according to Viral Tiling theory \cite{twarock2004tiling}, and introduce the graph modelling its interaction network. We then compute the fragmentation threshold at which the particle breaks into two disjoint components both under the removal of protein units, and as a consequence of bond breakage. The result is shown over a three-dimensional landscape, representing the three distinct types of bonds that occur in the capsid. Comparison with the Caspar Klug geometry corresponding to the special case that all bonds have equal strength, sheds new light on the possible evolutionary driving forces underpinning non-quasiequivalent viral architectures. 

\section{The Structure of Papillomavirus in Viral Tiling Theory} \label{section 1}

\begin{figure*}
    \centering
    \begin{subfigure}{.3\textwidth}
        \centering
        \includegraphics[width=\textwidth]{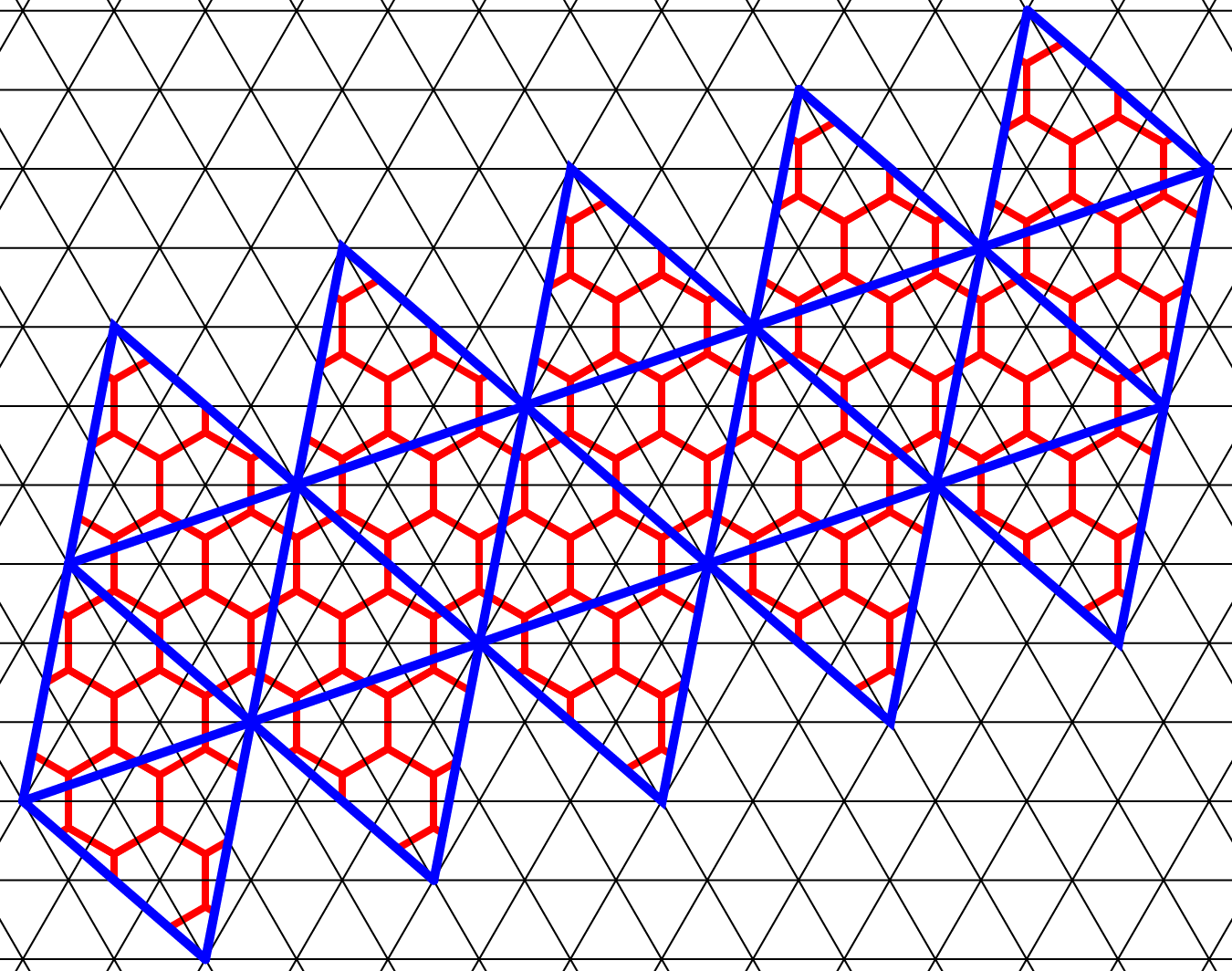}
        \caption{}
        \label{fig:figure1_tiling}
    \end{subfigure}
    \quad
       \begin{subfigure}{0.22\textwidth}
        \includegraphics[width=\textwidth]{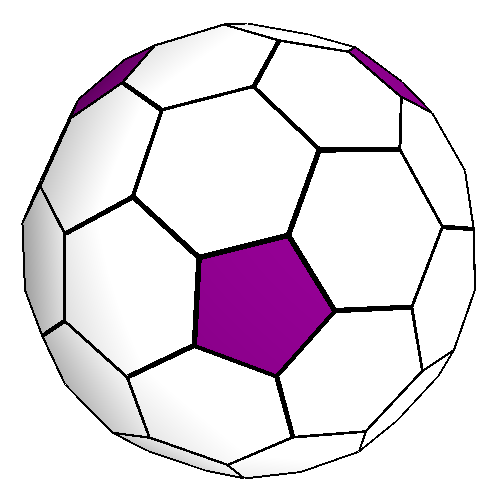}
        \caption{}
        \label{fig:figure2b}
    \end{subfigure}
   \quad
    \begin{subfigure}{0.23\textwidth}
        \includegraphics[width=\textwidth]{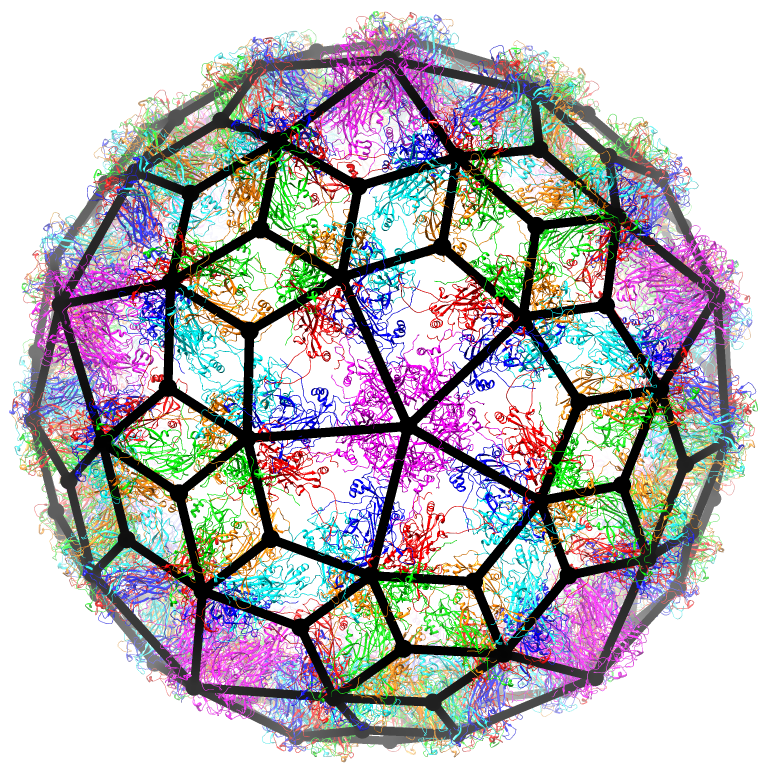}
        \caption{}
        \label{fig:figure2c}
    \end{subfigure}
    \hfill
    \begin{subfigure}{.225\textwidth}
        \includegraphics[width=\textwidth]{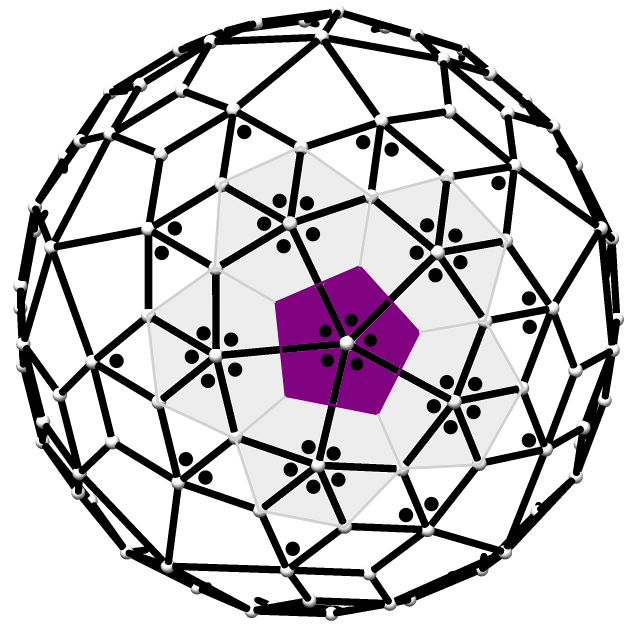}
        \caption{}
        \label{fig:figure1d}
    \end{subfigure}
    \quad
    \begin{subfigure}{0.3\textwidth}
        \begin{minipage}{\textwidth}
          \centering
          \raisebox{-0.5\height}{ \includegraphics[width=.75\textwidth]{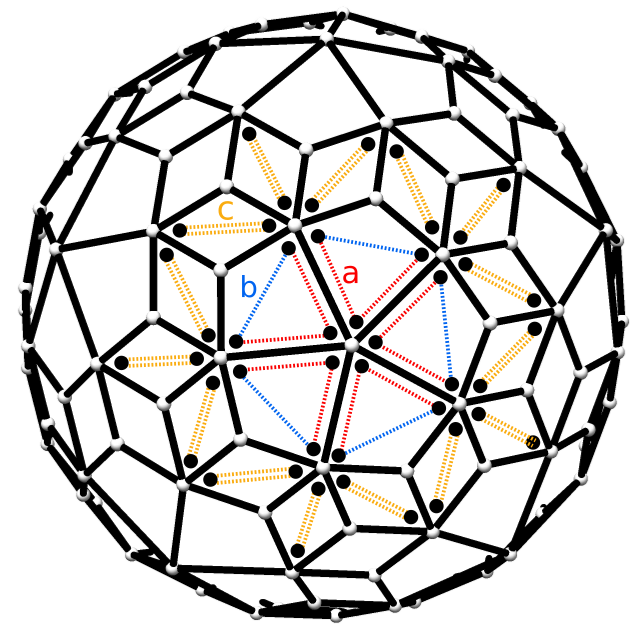}}
          \raisebox{-0.5\height}{\includegraphics[width=.2\textwidth]{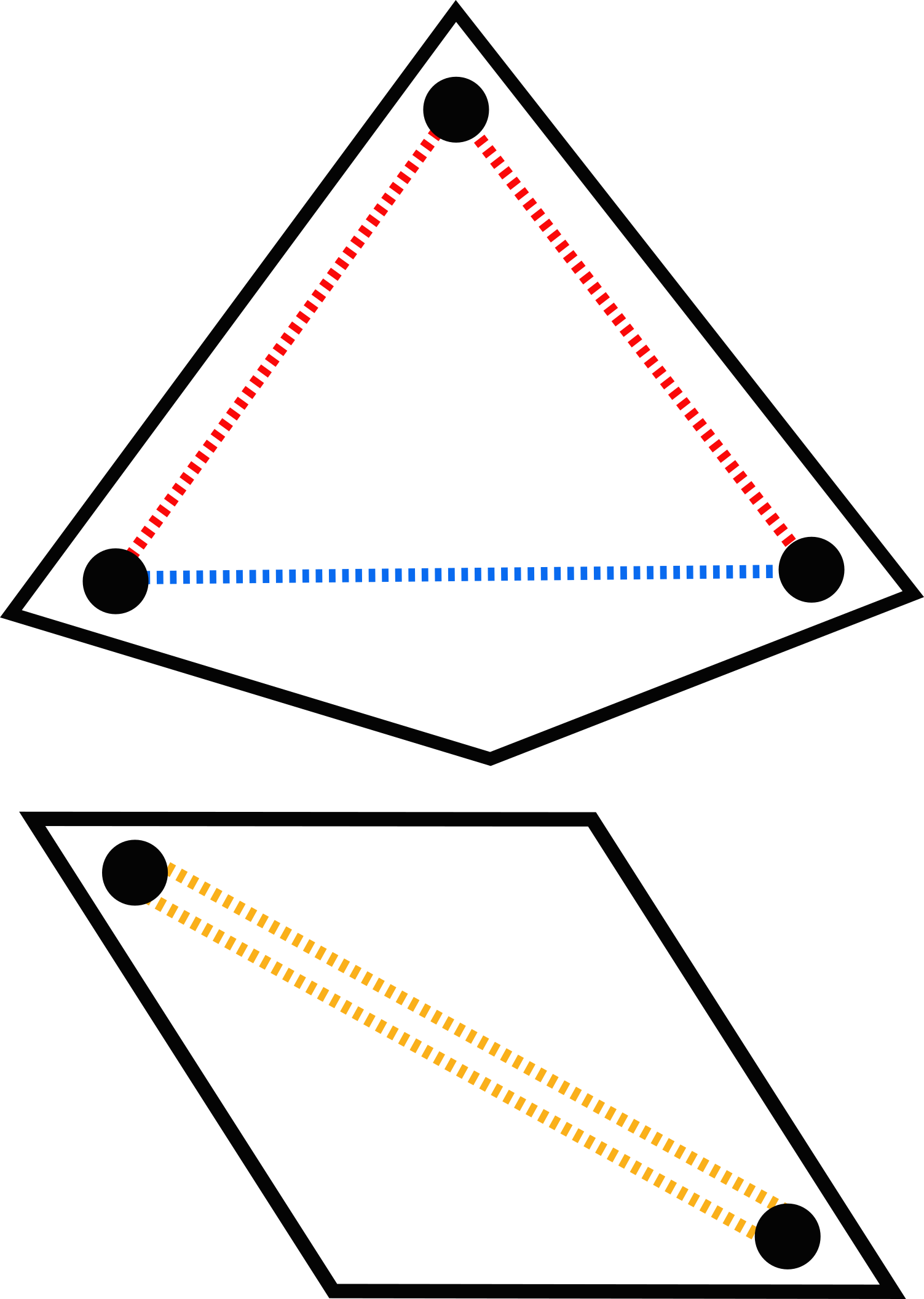}}
        \end{minipage}
        \caption{}
        \label{fig:figure1_bondcolours}
    \end{subfigure}
    \quad
    \begin{subfigure}{0.225\textwidth}
        \includegraphics[width=\textwidth]{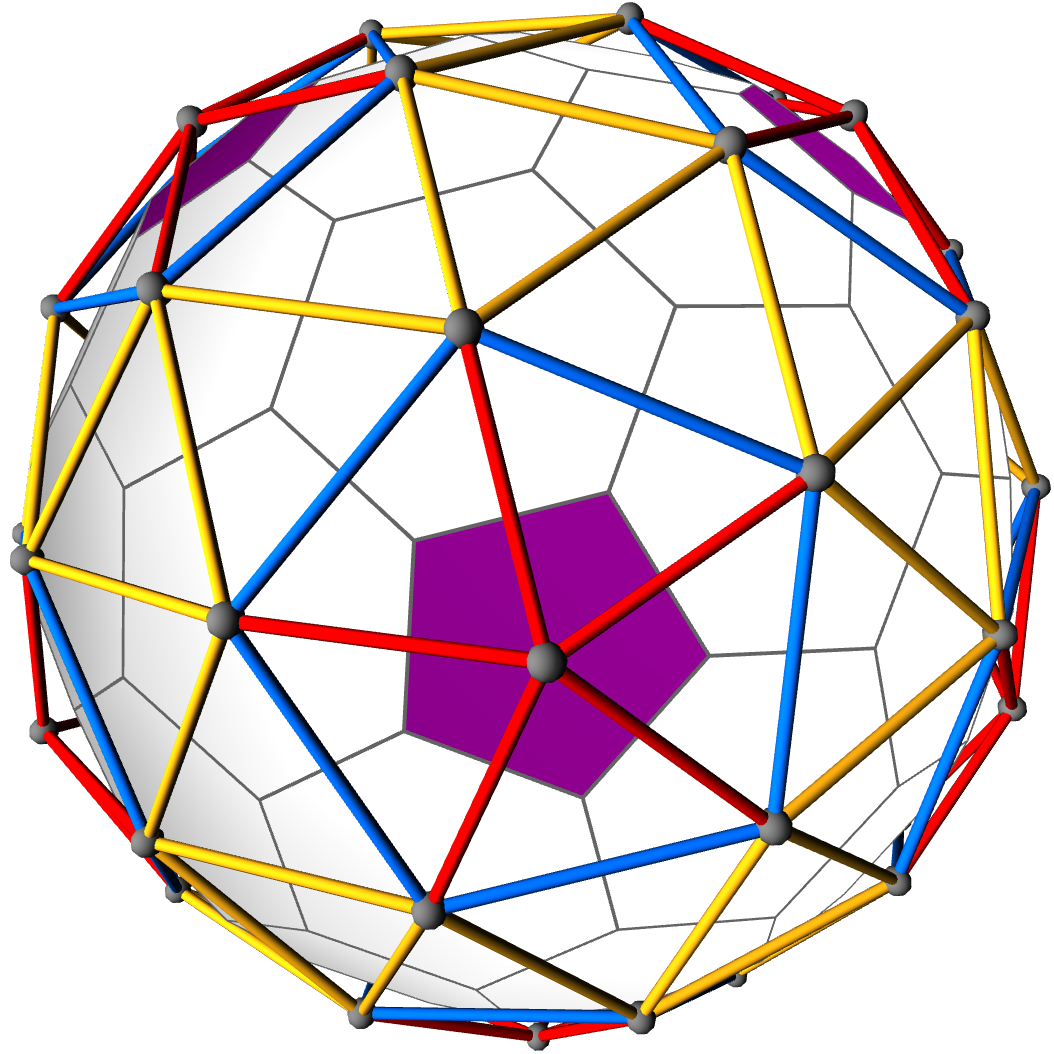}
        \caption{}
        \label{fig:figure1f}
    \end{subfigure}

    \caption{Geometric models of virus architecture. (a) The Caspar-Klug surface lattice of a $T=7$ capsid architecture. (b) It provides the layout for the spherical architecture shown as a polyhedron formed from 12 pentagons (magenta) and 60 hexagons. (c) The surface organisation of the papilloma capsid is shown  with tiling superimposed. (d) Pentamer positions in the papilloma virus tiling coincide with pentamers and hexamers (grey) of the $T=7$ Caspar-Klug virus architecture. (e) Kite and rhomb tiles represent three types of interactions, that are mediated by C-terminal arm extensions: type $a$ in red, type $b$ in blue, type $c$ in orange. (f) The weighted interaction network (wIN) of the papilloma virus capsid is obtained by placing vertices at the centres of the pentamers, and edges between pentamers that are connected via the interactions shown in (b); colours (weights) refer to the three distinct types of bonds. Geometric representations have been rendered using a \href{https://quentinrsl.github.io/polyomavirus-bonds-animation/}{purpose-designed software}.}
    \label{fig:figure1}
\end{figure*}

Caspar-Klug theory models virus architecture in terms of triangulations \cite{caspar1962physical} that indicate the positions of the capsid proteins (CPs) in the capsid surface. Geometrically distinct cage architectures are labelled by the triangulation number $T$, and correspond to different planar embeddings of the icosahedral surface into a hexagonal lattice (Fig. \ref{fig:figure1_tiling}). By construction, Caspar-Klug capsid architectures are formed from $60 T$ CPs that are organised as 12 pentagonal, and $10(T-1)$ hexagonal protein clusters, called pentamers and hexamers, respectively.

Papillomavirus capsids are formed from 72 pentamers and therefore cannot be modelled using the Caspar-Klug construction. Such capsid architectures are not quasi-equivalent in the sense of Caspar and Klug, because their CPs (indicated schematically by dots) are involved in two distinct types of interactions, mediated by C-terminal arm extensions, with neighbouring pentamers: dimers interactions between two protein subunits, and trimer interactions between three. 
Viral Tiling Theory models the surface architectures of these non-quasiequivalent viral capsids in terms of different types of tiles, that each represent a distinct interaction type  \cite{twarock2004tiling,twarock2005architecture,twarock2019structural}: rhombi representing dimer, and kites trimer interactions. 

Note that the centres of the pentamers in the papillomavirus tiling coincide with those of the pentamers and hexamers in a $T=7$ Caspar-Klug structure (compare Figs. \ref{fig:figure2b} \& \ref{fig:figure2c}). However, in contrast to the Caspar-Klug geometry, this capsid is formed from only 360 proteins (dots in Fig. \ref{fig:figure1d}), a number that is not possible in the framework of the Caspar-Klug construction. 

There are three distinct types of bonds between pentamers in the papilloma capsid: a bond corresponding to two C-terminal arms connecting a pair of proteins in a pentamer with a pair in a neighbouring pentamer (type $a$, red); a single C-terminal arm on a kite tile connecting two individual capsid proteins (type $b$, blue); and a dimer interaction, represented by a rhomb tile, with two C-terminal arms between two individual proteins (type $c$, yellow) (Fig. \ref{fig:figure1_bondcolours}). In particular, a type $a$ bond corresponds to two C-terminal arms between two pairs of proteins along the shared edge of two kite-shaped tiles. Type $b$ refers to the bond between the two proteins on a kite-shaped tile that are not involved in a type $a$ interaction with each other. Type $c$ bonds correspond to the bonds between the two proteins of a rhombic tile.

\section{A percolation theory model of virus disassembly for weighted interaction networks}

In this section, we introduce a percolation theory model for the disassembly of weighted interaction networks. The procedure broadly follows previous work for quasiequivalent capsid archtectures \cite{brunk2018molecular,brunk2021percolation}. However, as the network has different weights reflecting different types of bonds in the capsid, we modify the method to account for differences in the bond strengths. We start by formally introducing the weighted interaction network, and then present our method for both pentamer and bond removal scenarios.

\subsection{The weighted interaction network}

A prerequisite for modelling capsid disassembly is to encode the structural information in Fig. \ref{fig:figure1_bondcolours} as an interaction network, which captures topological information regarding the locations of  the assembly units (capsomers) and the interactions between them. The interaction network is represented as a graph, in which pentamers are represented as vertices, and interactions between pentamers as edges. In the case of non-quasiequivalent capsid architectures, such as the papillomavirus capsid considered here as an example, it is a weighted interaction network (wIN), in which edges are labelled according to different bond strengths. For the papillomavirus wIN, different weights are indicated by colours (Fig. \ref{fig:figure1f}) matching the three interaction types $a$, $b$, and $c$ in Fig. \ref{fig:figure1_bondcolours}.

In the following, we will investigate the propensity of the network to fragment when pentamers (vertices) or interactions (edges) are randomly removed from the wIN. We therefore attribute a weight to each edge that reflects the energy required to break that bond. The energies associated with type $a$, $b$, and $c$ bonds (shown in red, blue and yellow respectively in Fig. \ref{fig:figure1_bondcolours}) will be referred to as $E_a$, $E_b$ and $E_c$. Since proteins of rhombic tiles are involved in dimer interactions, whereas proteins of kite-shaped tiles are involved in the weaker trimer interactions, the corresponding bonds have different strengths. 
In particular, type $a$ bonds correspond to two C-terminal arm extensions in a trimer (two red lines), while type $b$ bonds is associated with a single C-terminal arm (blue line). Therefore, red edges in the interaction network have about double the bond energy of the blue edges. Moreover, type $c$ bonds correspond to a dimer interaction that is mediated by two C-terminal arm extensions. As yellow and red edges in the interaction network are both mediated by two C-terminal arm extensions, we assume that they are roughly equal. However, the dimer interactions are likely a bit stronger than two C-terminal arms in neighbouring trimer interactions. Therefore, we assume the following relations between the bond energies $E_a$, $E_b$ and $E_c$;  
\begin{align}\label{eq:1.1}
2E_b=E_a<E_c \,, 
\end{align}
where the difference between $E_a$ and $E_c$ is assumed to be not very large. Note that in this case the 12 pentamers at the particle 5-fold axes and the 60 additional pentamers, all have a similar energy in the capsid as $5 E_a = E_a + 2 E_b + 3 E_c$. This reflects the fact that they all interact with neighbouring pentamers via five C-terminal arm extensions. 

\subsection{Models of capsid disassembly}

We consider two distinct ways of modelling virus disassembly: either by removing bonds, or by removing vertices (i.e. pentamers) in the graph in the wIN. Both methods have been implemented before for the quasiequivalent capsid architectures in Caspar-Klug theory \cite{brunk2018molecular} and its extensions in the framework of Archimedean lattices \cite{brunk2021percolation}. In the computation of the fragmentation threshold of the viral capsids under bond removal, all bond energies had been assumed to be equal, so that bonds were broken randomly with a fixed known probability. We introduce below an approach that takes weighting of the edges according to their bond strengths into account.

\subsubsection{Capsid fragmentation under bond breakage} 

As the papillomavirus capsid has three distinct types of bonds with different bond energies, we associate with each pentamer an energy that is equal to the sum of the energies of its bonds to neighbouring pentamers. Each of the 12 pentamers at the particle 5-fold axes therefore has energy $5E_a$, and the 60 other pentamers are associated with energy $E_a + 2E_b + 3E_c$. The total energy of the viral capsid is therefore 
\begin{align}\label{eq:total_energy}
E=60E_a + 60E_b + 90E_c \,.
\end{align}
Since the energy needed to break a bond is different for each type of bond, it is reasonable to assume that bonds are not being removed in an equal manner. In order to account for this, each bond is given a probability weight which is inversely proportional to its bond energy. The process of bond removal applied in previous publications therefore has to be adapted. Instead of removing a certain fraction of bonds, we chose to remove a certain fraction $E_r$ ($r$ denoting removal) of the total capsid energy $E$. To do so, we pick a bond in a random manner (the probability for a bond to be chosen is  directly proportional to its probability weight) and check if there is enough energy to break the bond, i.e. if $E_r>E_i$ where $E_i$ is the bond energy under consideration. If so, we remove the bond and subtract its energy from $E_r$. We continue the process until no bond can be removed as the leftover energy is insufficient to do so. 

We then test the connectivity of the graph: if there are two or more isolated subgraphs, the graph is considered to be fragmented. 
This process is repeated a sufficient number of times to obtain a value for the probability of graph fragmentation, depending on the energy of bonds removed $E_r$, within a certain range of accuracy (see Methods). We then use the values obtained to find the energy fragmentation threshold, i.e. the fraction of energy that needs to be removed for the probability of graph fragmentation to be equal to $0.5$ using a classic bisection method.
For this, the outcome of the simulation is benchmarked against the fragmentation threshold curve. Chebychev's inequality is used to determine a condition on the number of iterations required for each step of the bisection process (see Methods). 

\subsubsection{Capsid fragmentation under pentamer removal}

As viral capsids in the papillomavirus family disassemble into pentamers, we also consider pentamer removal, which corresponds to removal of nodes, rather than edges, from the wIN. For this, we associate with each node a probability weight that is directly proportional to the pentamer's total bond energy, as defined above. Then, in analogy to the procedure for edge removal, given a fraction of energy $E_r$ to remove, we remove nodes and their associated edges until we cannot do so as there are no nodes of the appropriate energy remaining in the wIN. As nodes are removed, some bonds that were previously connected to neighbouring nodes are now broken, thus reducing the energy of the remaining nodes. We have therefore included a routine into our simulations that updates the energy of any remaining nodes, and consequently their probability weights, after a node has been removed from the wIN. By repeating this fragmentation process, we obtain a value for the probability of graph fragmentation depending on the fraction of energy removed ($E_r$), but this time in terms of pentamer/node removal, rather than bond removal. 

\section{Results}

\subsection{Stability of quasiequivalent versus non-quasiequivalent capsid architectures}

We implemented the above described methods of edge and node removal to the paplillomavirus wIN in Fig. \ref{fig:figure1f}. 
\begin{figure*}
        \centering
    \begin{subfigure}{0.3\textwidth}
        \centering
        \includegraphics[width=\textwidth]{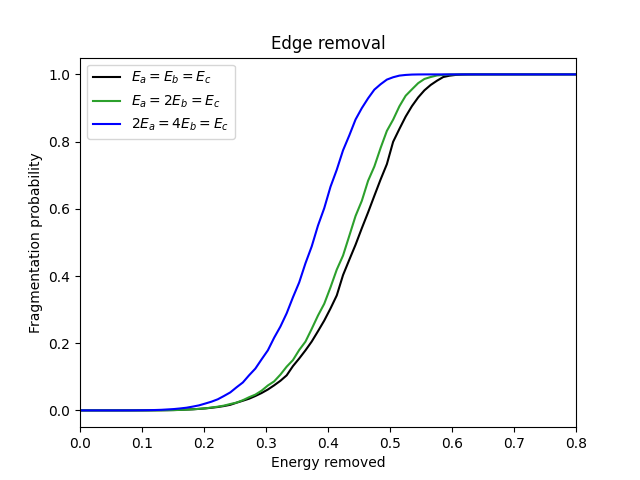}
        \caption{}
        \label{fig:frag_prob_polyoma_edges}
    \end{subfigure}
    \begin{subfigure}{0.3\textwidth}
        \includegraphics[width=\textwidth]{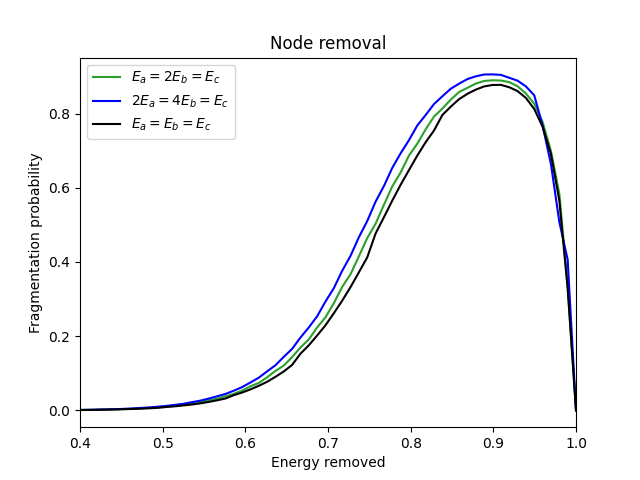}
        \centering
        \caption{}
        \label{fig:frag_prob_polyoma_nodes}
    \end{subfigure}

    \begin{subfigure}{0.35\textwidth}
        \centering
        \includegraphics[width=\textwidth]{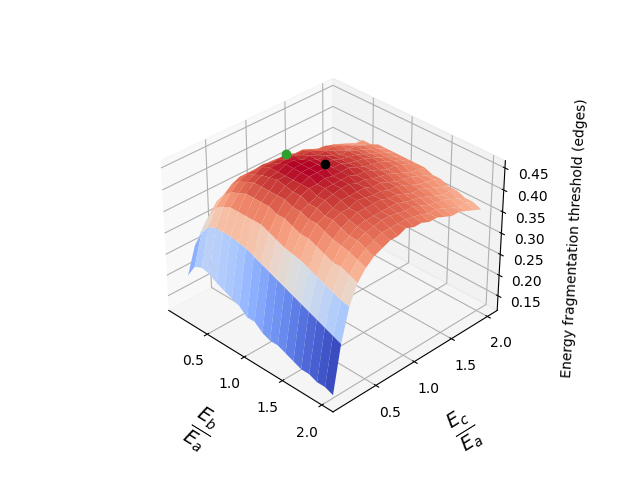}
        \caption{}
        \label{fig:landscape_polyoma_edges}
    \end{subfigure}
    \begin{subfigure}{.35\textwidth}
        \centering
        \includegraphics[width=\textwidth]{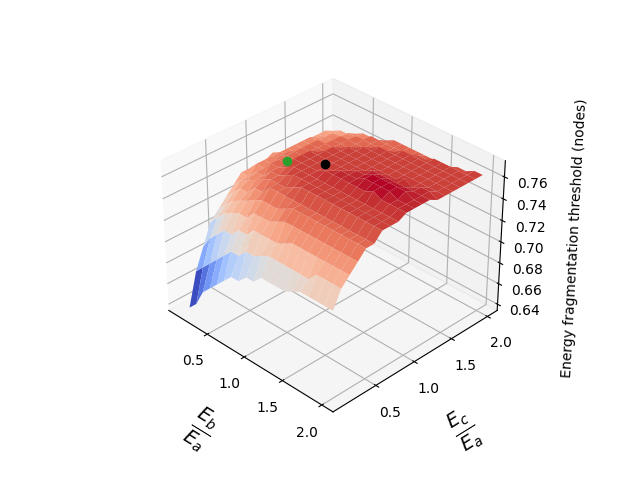}
        \caption{}
        \label{fig:landscape_polyoma_nodes}
    \end{subfigure}
    \caption{Resilience to fragmentation of the papillomavirus wIN under bond breakage and pentamer removal. Probability of fragmentation for various bond strengths (choices of weights in the wIN) under (a) edge and (b) node removal. The energy percolation threshold for edge (c) and node (d) removal for different combinations of bond strengths reveals the Caspar-Klug (CK) scenario ($E_a=E_b=Ec$, black dot) and the papilomavirus (P) scenario ($E_a=2E_b=Ec$, green dot) to be located in the more stable range (red values in the landscape), with the CK geometry being more resilient to fragmentation than the P architecture. }
    \label{fig:landscape}
\end{figure*}
The results depend on the relative values of the three bond strengths $E_a$, $E_b$ and $E_c$ (Fig. \ref{fig:landscape}). Equal weights ($E_a=E_b=Ec$, black) represent the quasiequivalent interaction network of a $T=7$ Caspar-Klug (CK) geometry, and $E_a=2E_b=Ec$ (green) the non-quasiequivalent papillomavirus (P) scenario. Both are more resilient to fragmentation than most other scenarios (e.g. $2E_a=4E_b=Ec$, blue), albeit with CK being slightly more resilient than P (note the displacement of the black line to the right of the green curve). The positions of these scenarios in the energy landscape are indicated by black and green dots, respectively. These results suggest that viruses have evolved geometries that confer more stability to the capsid than most alternatives. They also reveal how protein container architectures might be designed in virus nanotechnology, by configuring bond energies appropriately, to achieve less stable cage architectures if desired. 

We note that the probability of fragmentation in Fig. \ref{fig:frag_prob_polyoma_nodes} tends to 0 as the fraction of energy removed approaches 1. This is a consequence of our model set-up. In contrast to previous methods, the energy of neighbouring nodes decreases when a node is removed, reflecting the absence of broken bonds. Therefore, the probability weights of such nodes increase and they are more likely to be chosen, consistent with expectations. The larger the fraction of the total energy removed, the larger the number of nodes removed. As a result, it is likely that the subgraph obtained after removal of a large fraction of the total energy, is composed of only a small number of connected nodes. Such graphs are naturally connected, leading to a decreasing probability of fragmentation. However, at that stage, the remaining graph is so small that the cargo has already been released, so this does not pose any problem for the biological conclusions from this work. 

\subsection{Comparing hole formation with capsid fragmentation}

Before the capsid fragments into two disjoint parts, it is possible that a hole can form via removal of individual pentamers that is large enough to enable cargo release before capsid fragmentation is taking place. We therefore study here the process of hole formation, and investigate whether the formation of a large hole occurs prior or post capsid fragmentation for different wINs. 
\begin{figure}[H]   
   \centering
   \begin{subfigure}{.6\columnwidth}
        \includegraphics[width=\linewidth]{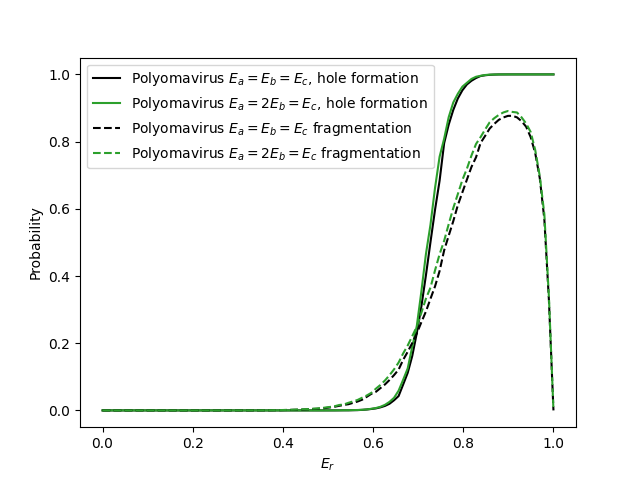}
        \caption{}
        \label{fig:comparaison_polyoma}
    \end{subfigure}
    
    \begin{subfigure}{.6\columnwidth}
        \includegraphics[width=\linewidth]{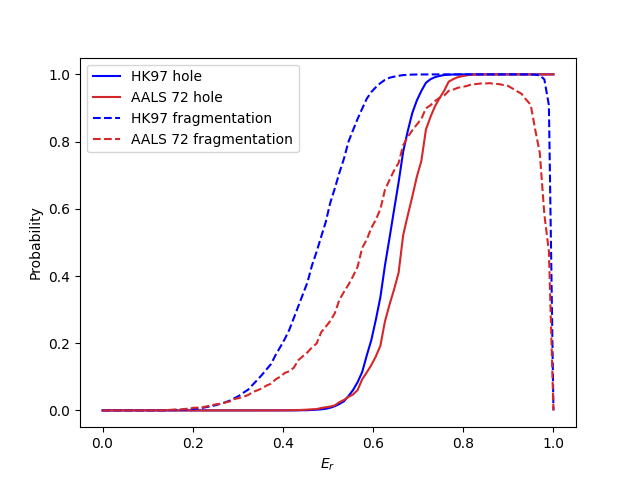}
        \caption{}
        \label{fig:comparaison_other}
    \end{subfigure}

    \begin{subfigure}{.18\columnwidth}
        \includegraphics[width=\linewidth]{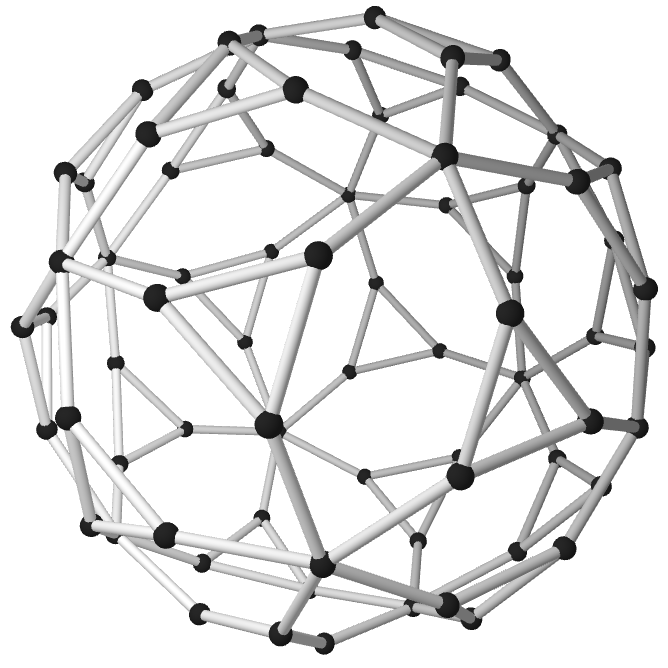}
        \caption{}
    \end{subfigure}
    \qquad
    \begin{subfigure}{.18\columnwidth}
            \includegraphics[width=\linewidth]{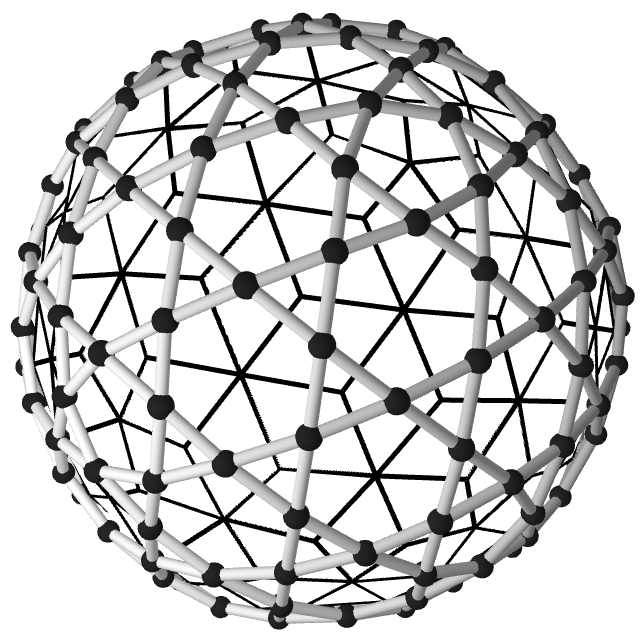}
        \caption{}
    \end{subfigure}
    \caption{Comparison of hole formation with capsid fragmentation. 
    (a) Fragmentation probabilities (dashed curves) are compared with the probability of having a hole of size larger than half the capsid (solide curve) for different edge energy distributions in the papilloma wIN; in all cases, hole formation occurs prior to capsid fragmentation. (b) Curves corresponding to the interaction network of the AaLS cage made of 72 pentamers, shown in (c), and 
    the HK97 virus, shown in (d), show the opposite trend, with capsid fragementation (dashed lines) occurring prior to hole formation (solid curve).}  
    \label{fig:comparaison_hole_proba}
\end{figure}

For this we compute the probability that the size of the largest hole in the capsid is larger than half of the capsid. We compare the "removal" energy $E_r$ for which this probability surpasses 0.5, a proxy for the transition from small to large hole sizes, with the fragmentation probability, see Fig. \ref{fig:comparaison_hole_proba}.
Interestingly, the papillomavirus wIN exhibits a different behaviour from that of other protein cages of similar size: a non-quasiequivalent \textit{de novo} designed protein cage (AaLs, shown in (b)), and a quasiequivalent $T=7$ viral cage  (HK97, (d)) formed from rhombic building blocks. Whilst hole formation occurs prior to capsid fragmentation in the papillomavirus architecture, the opposite is the case for the other cages. This hints at a principally different disassembly mechanism in the papillomaviruses.

\begin{figure*}
    \centering
    \begin{subfigure}{0.4\linewidth}
        \includegraphics[width=\textwidth]{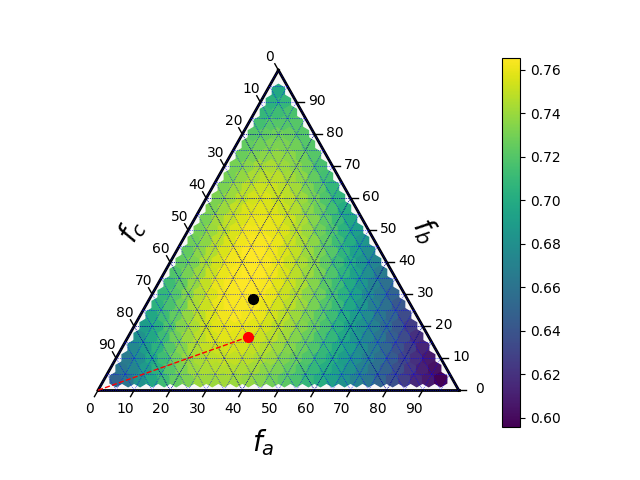}
        \caption{$E_F$ for node removal}
    \end{subfigure}
    \begin{subfigure}{0.4\linewidth}
        \includegraphics[width=\textwidth]{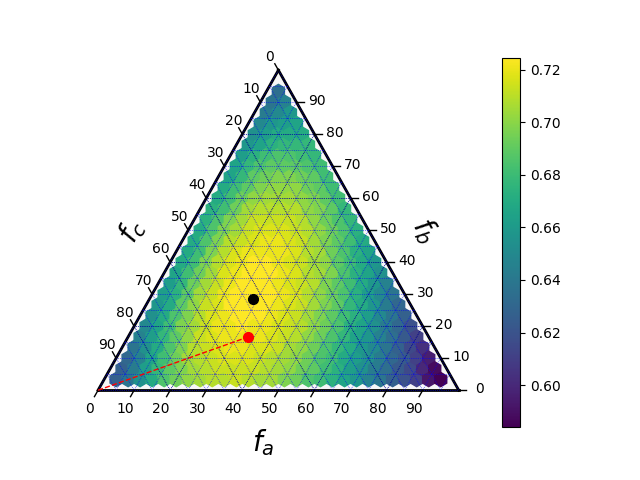}
        \caption{$E_H$ for node removal}
    \end{subfigure}

    \begin{subfigure}{0.4\linewidth}
        \includegraphics[width=\textwidth]{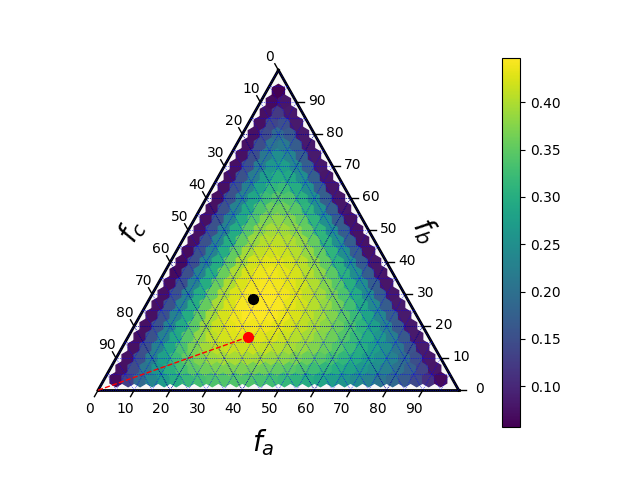}
        \caption{$E_F$ for edge removal}
    \end{subfigure}
    \begin{subfigure}{0.4\linewidth}
        \includegraphics[width=\textwidth]{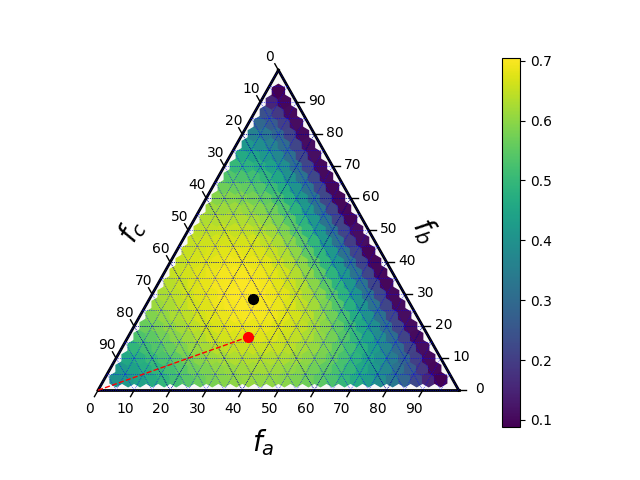}
        \caption{$E_H$ for edge removal}
    \end{subfigure}
    \caption{Comparison of fragmentation and hole formation. (a) Ternary graphs\cite{pythonternary} for fragmentation energy $E_F$ (left) and energy of hole formation $E_H$ (right) when removing nodes/pentamers (top row) or edges/bonds (bottom row) from the interaction network/capsid. The black dot indicates the CK scenario of identical bond strengths, while the red dot corresponds to the papilloma scenario $E_a=2E_b=E_c$. The red lines indicates  parameters with $E_a=2E_b$ and $E_c>E_a$. The actual value for the papillomavirus capsid is on this line and close to the red dot.}  
    \label{fig:CKResults}
\end{figure*}

This conclusion is further supported by ternary graphs comparing the energies $E_F$ at which fragmentation occurs, with $E_H$ when hole formation occurs, for both node removal (Fig. \ref{fig:landscape}, top row) and edge removal (bottom row).  
Denoting as $f_a$, $f_b$ and $f_c$ different fractions associated with each type of bond in the total capsid energy, i.e. 
\begin{equation}
\begin{split}\label{eq:fraction_energy_relation}
    f_a&= \frac{60E_a}{E} = \frac{E_a}{E_a + E_b + \frac{3}{2}E_c}\\
    f_b&= \frac{E_b}{E_a + E_b + \frac{3}{2} E_c}\\
    f_c&= \frac{E_c}{\frac{2}{3} E_a + \frac{2}{3} E_b + E_c}
\end{split}
\end{equation}
we plot the energy fragmentation threshold for different energy distributions in Fig. \ref{fig:CKResults}.

Using the relations (\ref{eq:1.1}) and (\ref{eq:fraction_energy_relation}), we deduce the following conditions for $f_a$, $f_b$ and $f_c$:
\begin{align}\label{eq:2.1}
\begin{split}
f_a&=2f_b\\
f_c&>\frac{3}{2}f_a
\end{split}
\end{align}
These relations define the red line in the ternary graph: it connects the point $(f_a=0,f_b=0,f_c=1)$, corresponding to bond energies $E_a=E_b=0$, with $(f_a=2/6, f_b=1/6, f_c=3/6)$, which corresponds to the ideal scenario of bond strength $E_a=2E_b=E_c$. The realistic value will be in the vicinity to this line close to the ideal value (red dot). Note that this is in the region corresponding to higher fragmentation energies, indicating capsid structures that are more resilient to fragmentation. 

It is interesting to compare the ternary graphs for node and edge removal. Whilst graphs for $E_F$ and $E_H$ are similar for the node removal case, they differ markedly for edge removal. The capsid now opens a hole before fragmentation (on average $\frac{E_H}{E_F}=1.71$). This difference is particularly pronounced for capsids with weak $a$ bonds: their resistance to fragmentation diminishes rapidly to 0, in contrast to their resistance to hole formation. This makes sense as removal of $a$ bonds from the wIN results in "floating" nodes that fragment the graph. As those holes are only of size 1, this does not affect the largest hole size significantly.
Unlike $a$ bonds, $c$ bonds have a crucial role in the structure of the capsid, in term of resistance to both fragmentation and hole formation. This is consistent with the fact that $c$ bonds form a connected subgraph, 
which corresponds to a "whiffle ball" architecture \cite{Whiffle}, and the fact that they are the strongest bonds in the wIN. 





For comparison, the CK scenario of a $T=7$ capsid with equal bond strengths $E_a=E_b=E_c$ corresponds to: 
\begin{align}
f_a=f_b=\frac{2}{7} \\
f_c=\frac{3}{7}\,,
\end{align}
which is indicated by a black dot. In all graphs, the non-quasiequivalent geometry of the papilloma capsid is less resilient to fragmentation than its quasiequivalent counterpart. However, it is still relatively stable (yellow/green range), consistent  with its function to offer sufficient protection to its genetic material, while enabling its timely release when infecting its host.





\subsection{Analysis of disassembly pathways}

\begin{figure*}
    \centering
    \begin{subfigure}{.7\textwidth}
        \begin{subfigure}{.5\textwidth}
            \includegraphics[width=\textwidth]{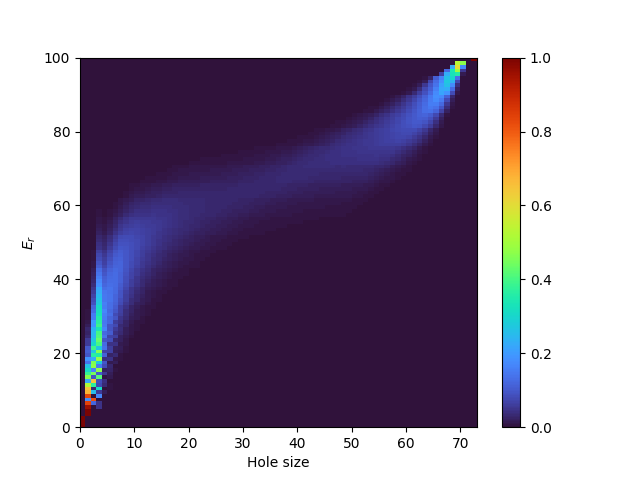}
            \caption{}
            \label{fig:aals_heatmap}
        \end{subfigure}
        \begin{subfigure}{.5\textwidth}
            \includegraphics[width=\textwidth]{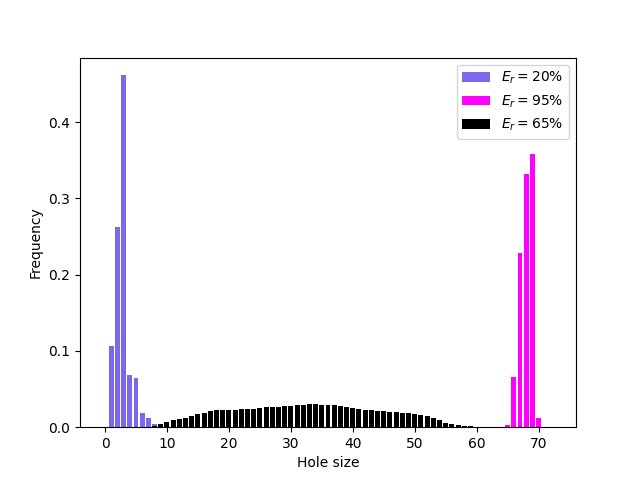}
            \caption{}
            \label{fig:aals_hist}
        \end{subfigure}
        
        \begin{subfigure}{.5\textwidth}
            \includegraphics[width=\textwidth]{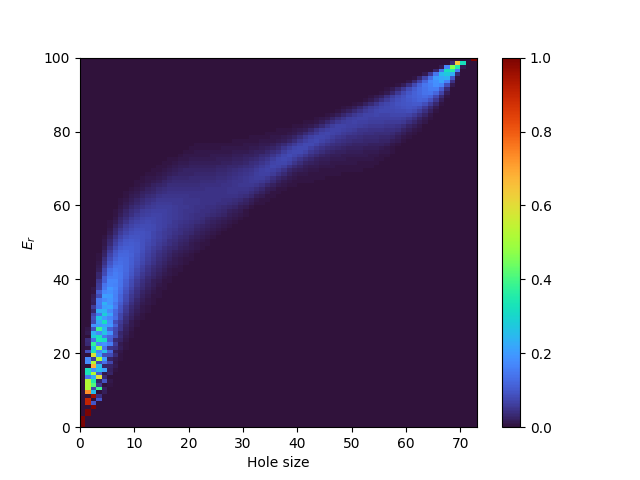}
            \caption{}
        \end{subfigure}
        \begin{subfigure}{.5\textwidth}
            \includegraphics[width=\textwidth]{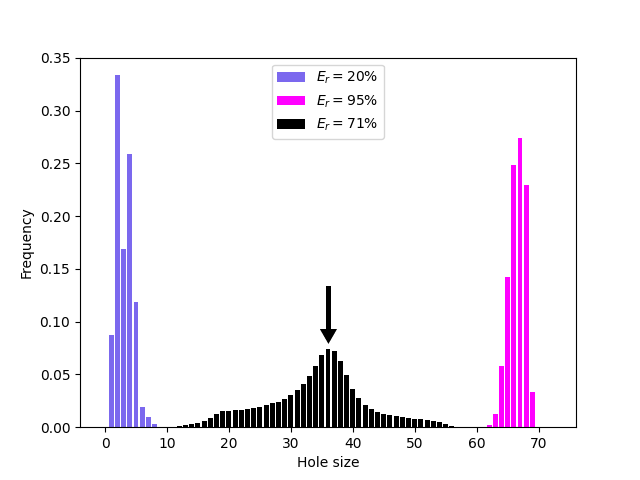}
            \caption{}
             \label{fig:polyoma_hist}
        \end{subfigure}
    \end{subfigure}
    \begin{subfigure}{.25\textwidth}
        \centering
        \includegraphics[width=\textwidth]{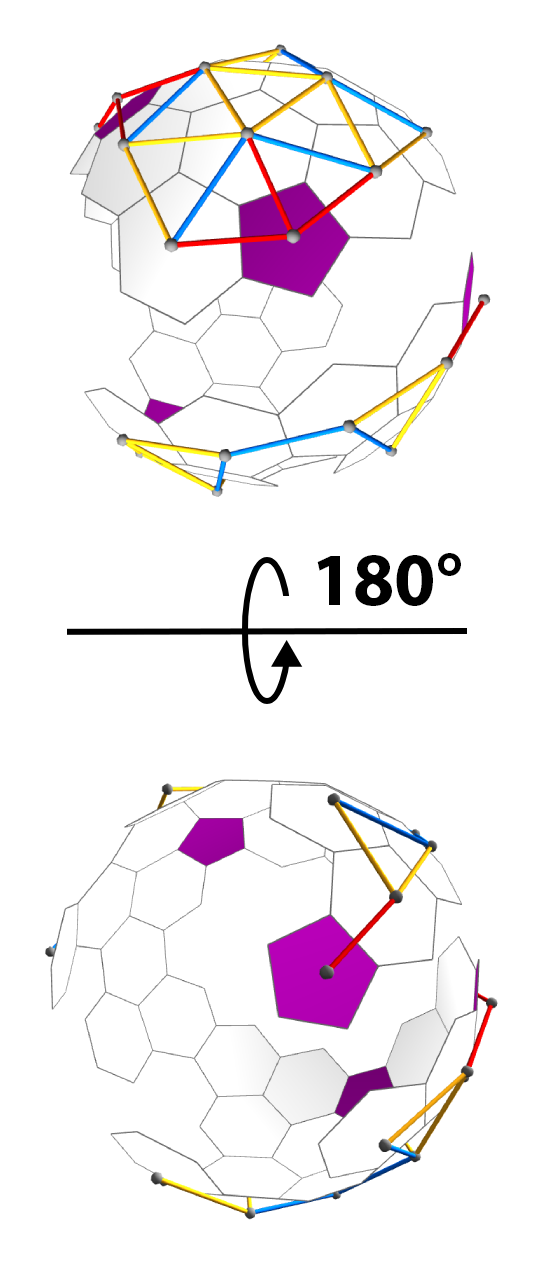}
        \caption{}
        \label{fig:polyoma_fragmented_36}
    \end{subfigure}   
    \caption{Hole size distribution during disassembly. (a) Hole size distributions in AaLS72 disassembly intermediates under node removal for different values of $E_r$. (b) Close up of the distributions for energies $E_r$ corresponding to 20\%, 65\% and 95 \% of the total capsid energy $E$, respectively. (c) \& (d) Equivalent data for the papillomavirus capsid graph with $E_a=E_c=2E_b$. (e) Example of the interaction network of a papilloma disassembly intermediate with a hole size of 36, corresponding to the peak of the distribution in (d), see arrow.}
    \label{fig:polyoma_sizeholedist}
\end{figure*}

As hole formation occurs prior to capsid fragmentation in papillomavirus according to Fig. \ref{fig:comparaison_polyoma}, we further analyse the process of hole formation. Fig.\ref{fig:polyoma_sizeholedist} shows the distribution of hole sizes for different values of the removal energy $E_r$. 
Up to a certain threshold of energy removed ($E_r=E_H$), the holes in the capsid do not exceed a third of the capsid in size and the probability distribution retains a low standard deviation. Above that threshold, the size of the largest hole is consistently above 2/3 of capsid size. For fragmentation energies close to $E_H$ we observe a transition regime where the standard deviation increases and the average size of the largest hole rapidly increases. For the \textit{de novo} designed AaLS72 cage, no hole size is significantly favoured during this regime (see the flat distribution in black), i.e., no particular intermediary value is favoured for transitioning from a small to a large hole size (Fig. \ref{fig:aals_heatmap} and \ref{fig:aals_hist}). However, the papillomavirus capsid exhibits a peak for capsid intermediates with a hole size close to half the capsid size during this regime (see arrow in Fig. \ref{fig:polyoma_hist}). This can also be seen quantitatively by comparing the normalized entropies of the hole size distribution at $E_r=E_H$: This value is approximately $0.61$ for the AaLS72 capsid, but $0.56$ for the papillomavirus capsid ($E_a=E_c=2E_b$). The maximal peak height over the average peak height is 5.39 for the papillomavirus wIN, but only 1.98 for the AaLS72 cage. Interestingly, a similar distribution (and indeed the same entropy value of 5.36) occurs also for the unweighted interaction network, i.e. for the $T=7$ CK architecture. This shows that the papillomavirus capsid and CK geometry structurally favour an intermediary state during disassembly in which the capsid is missing half of the pentamers. An example of a  capsid intermediate with a hole size of 36 is shown in Fig. \ref{fig:polyoma_fragmented_36}. 

\subsection{De novo designed versus natural protein cages}

\begin{figure*}
   \begin{subfigure}{0.17\linewidth}
       \includegraphics[width=\linewidth]{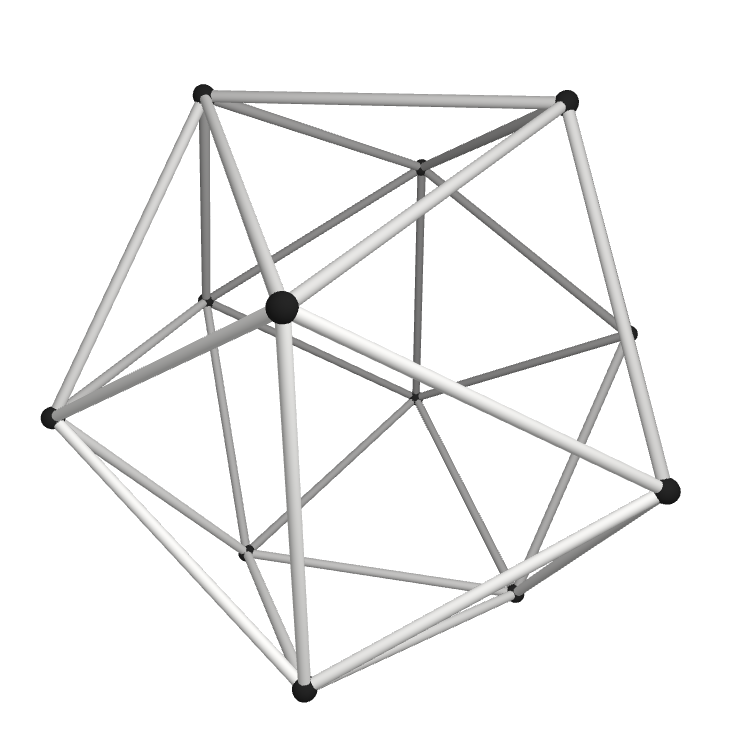}
       \includegraphics[width=\linewidth]{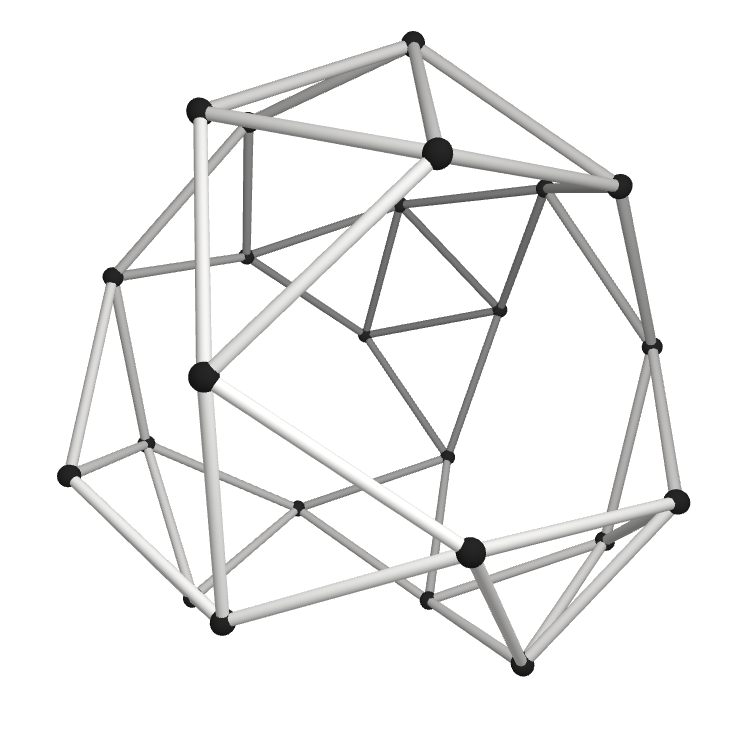}
       \includegraphics[width=\linewidth]{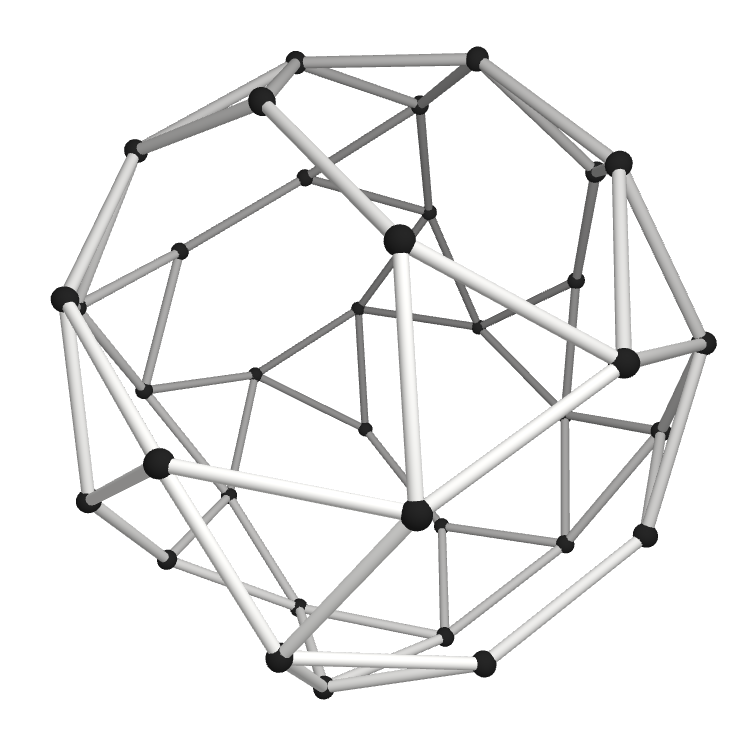}
       \includegraphics[width=\linewidth]{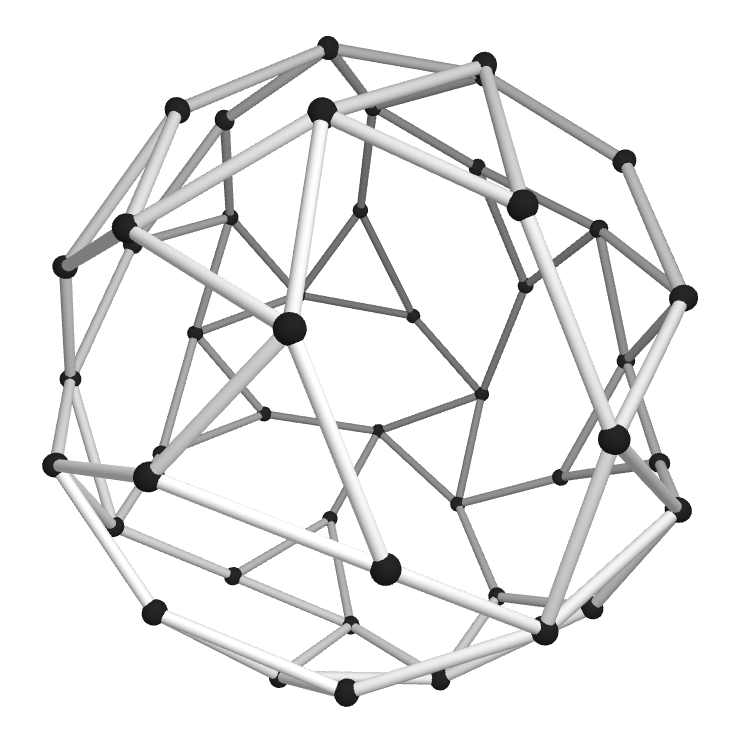}
       \includegraphics[width=\linewidth]{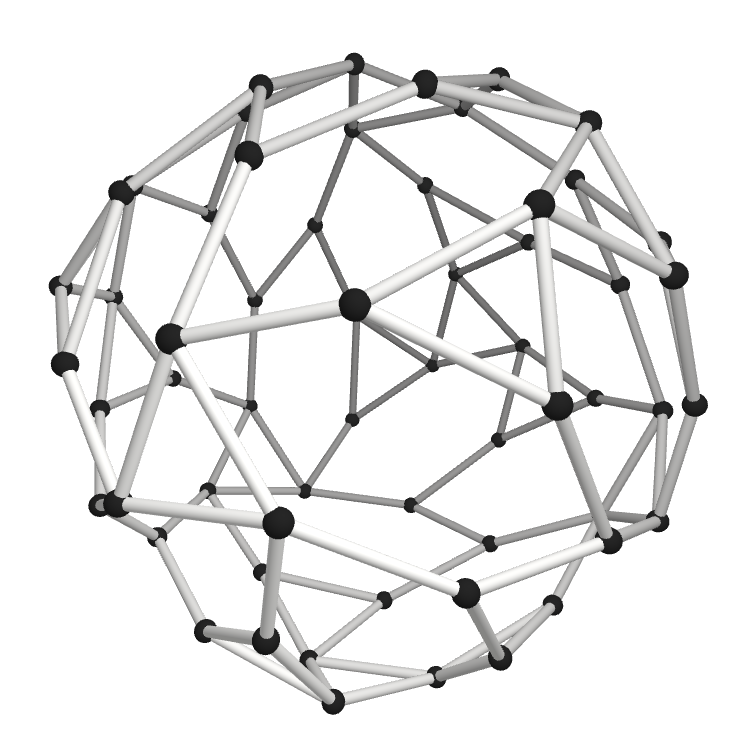}
       \includegraphics[width=\linewidth]{aals72.png}
       \caption{}
       \label{fig:aals_cages}
   \end{subfigure}
   \begin{subfigure}{.83\linewidth}
       \begin{subfigure}{.5\linewidth}
           \includegraphics[width=\linewidth]{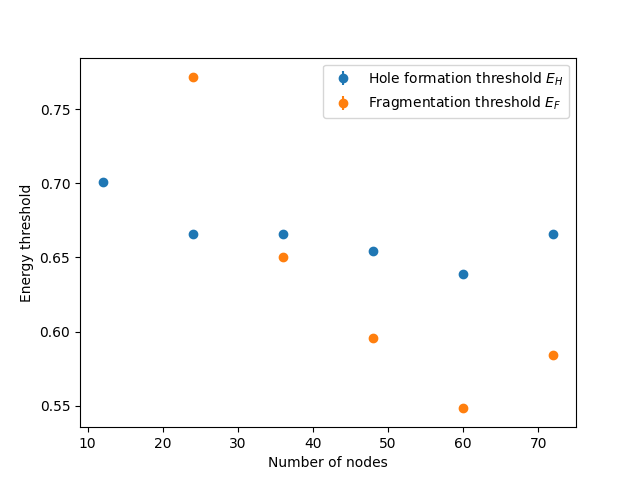}
           \caption{}
           \label{fig:aals_threshold}
       \end{subfigure}
       \begin{subfigure}{.5\linewidth}
           \includegraphics[width=\linewidth]{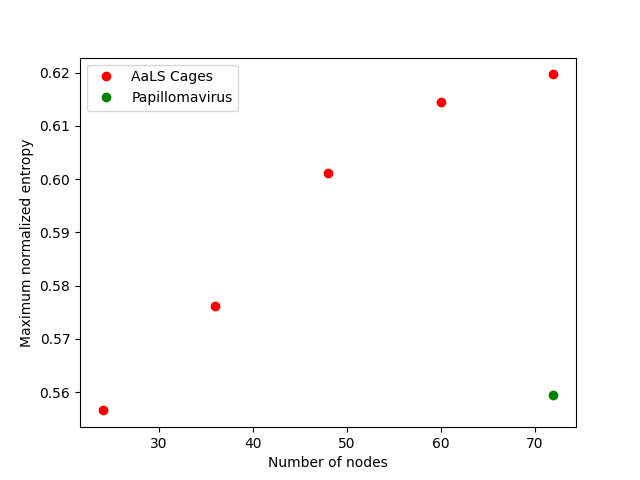}
           \caption{}
           \label{fig:aals_entropy}
       \end{subfigure}
       
       \begin{subfigure}{.5\linewidth}
           \includegraphics[width=\linewidth]{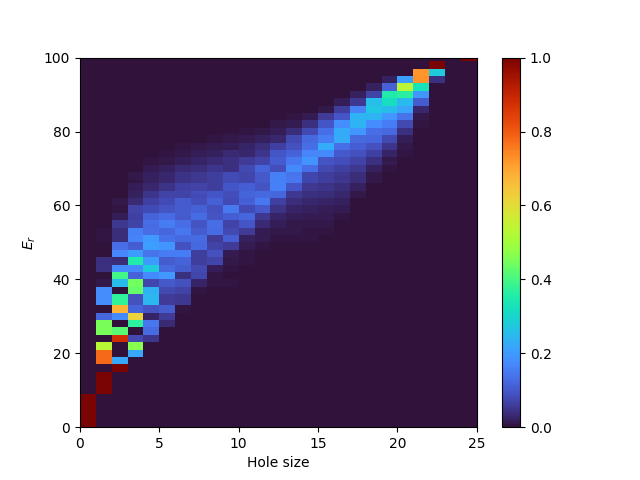}
           \caption{}
           \label{fig:aals_heatmap_24}
       \end{subfigure}
       \begin{subfigure}{.5\linewidth}
           \includegraphics[width=\linewidth]{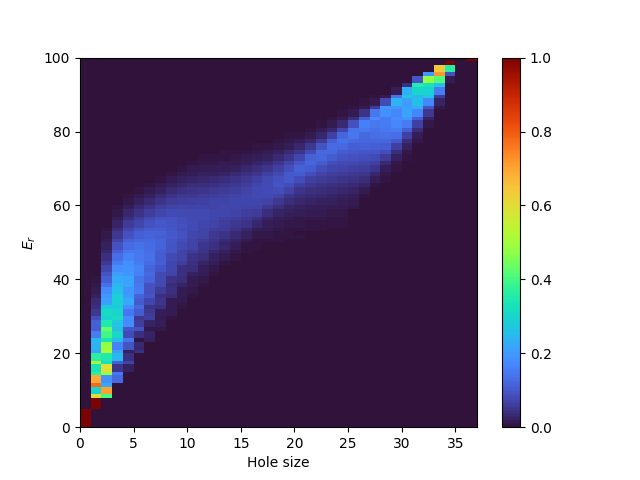}
           \caption{}
       \end{subfigure}
    
       \begin{subfigure}{.5\linewidth}
           \includegraphics[width=\linewidth]{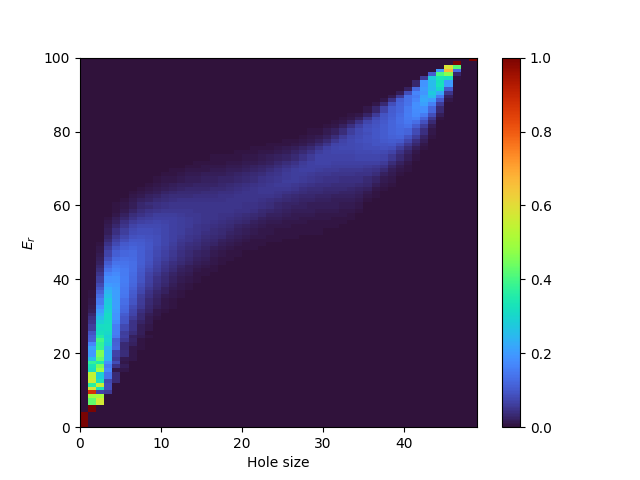}
           \caption{}
       \end{subfigure}
       \begin{subfigure}{.5\linewidth}
           \includegraphics[width=\linewidth]{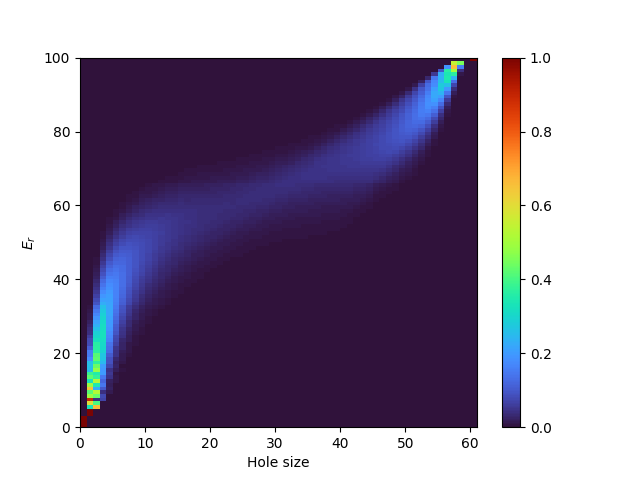}
           \caption{}
           \label{fig:aals_heatmap_72}
       \end{subfigure}
   \end{subfigure}
   \caption{Fragmentation behaviour of the \textit{de novo} designed AaLS cages. (a) The interaction networks of symmetric AaLs cage architectures, computed using a \href{https://quentinrsl.github.io/aals_cages/}{purpose-designed software}. (b) The hole formation thresholds $E_H$ and fragmentation thresholds $E_f$ for the different cage structures. Note the lack of a data point for the fragmentation threshold for the smallest cage as the fragmentation probability remains consistently under 0.5. (c) The AaLS cages exhibit increasing entropies with size in their distributions at the cusp between small and large holes (red), resulting in a significantly higher value than for the papillomavirus capsid (green). (d)-(g) Hole size distributions for disassembly intermediates for tetrahedral AaLS cages with with 24, 36, 48 and 60 pentamers.}
   \label{fig:aals}
\end{figure*}

The difference in disassembly behaviour between the \textit{de novo} designed AaLS72 cage and the virus examples is striking, and begs the question whether this phenomenon occurs more widely in \textit{de novo} designed cage architectures. The AaLS pentamer is known to assemble into a wide range of cage structures with distinct symmetries and shapes (Fig. \ref{fig:aals_cages}). 

Whilst the smallest and largest cage have icosahedral symmetry, the four intermediate-sized cages exhibit tetrahedral symmetry. Resilience to fragmentation drops rapidly amongst the tetrahedral cage architecture with increasing size. However, there is a gain in resilience in the transition from the tetrahedral 60-pentamer cage to the icosahedral 72-pentamer cage, suggesting that symmetry has an impact on stability (Fig. \ref{fig:aals_threshold}). A similar trend is observed for hole formation, but generally that curve is flatter, suggesting only limited variation in hole formation across the ensemble of AaLS cage architectures. There is a cross-over in the curves between the 24-pentamer and the 36-pentamer cage, making hole formation more likely in the smaller cages, and fragmentation more likely in the larger ones. 

This analysis reveals distinct assembly pathways for different capsid sizes. The maximal normalised entropy is increasing with cage size (Fig. \ref{fig:aals_entropy}), consistent with the individual hole size distributions for the tetrahedral intermediate-sized cages shown in Fig. \ref{fig:aals_heatmap_24}-\ref{fig:aals_heatmap_72}. 
These reveal a pattern similar to the icosahedral 72-pentamer AaLS cage in Fig. \ref{fig:aals_heatmap}. It is characterised by the absence of a defined pathway of hole formation for these architectures, in contrast to the papillomavirus case. This suggests that \textit{de novo} designed containers can exhibit disassembly behaviour that is principally different from that of naturally occuring cage structures. 

\section{Methods}
\subsection{Generation of the interaction network in 3D}
The following geometric approach was used to visualise capsid architectures and their interaction networks. Starting with a list of edges corresponding to a tile, this tile was translated along two given vectors $\overrightarrow{T_x}$ and $\overrightarrow{T_y}$ to generate the lattice grid. Then three 6-fold symmetry axes of the grid were chosen to indicate the vertices of an equilateral triangle. Only edges that intersect with, or are contained within, this triangle, were identified, effectively "cutting" this triangle out of the underlying planar lattice. The position of this triangle in the capsid surface was then defined by two integers $(h,k)$, where $(h\overrightarrow{T_x},k\overrightarrow{T_y})$ is the vector between two vertices of the triangle. This algorithm was used for a triangular tiling with $(h,k) = (2,1)$ to generate one of the twenty faces of the papillomavirus capsid. We then manually assign weights to the edges before copying this face twenty times. After assembling icosahedral faces in 3D, we obtain the graph of the viral capsid. A similar method has been used for the generation of the AaLS cages in Fig. \ref{fig:aals_cages} (see also \href{https://github.com/quentinrsl/capsidgraph}{Gihub}).



\subsection{Edge and node removal from a weighted graph} \label{m:bonds}
For edge/node removal from a weighted interaction network (wIN), we assign probability weights to each edge which are inversely proportional to their bond energy. Instead of working with a probability of removal, we pick an amount of "energy equivalent" $E_r$ to randomly remove from the wIN, that is typically indicated as a percentage of the total energy $E$. The Monte Carlo simulation is conducted as follows: We randomly choose bonds until we find one which has less energy than $E_r$. We remove this bond and subtract its energy from $E_r$. We repeat this process until all bonds have more energy than $E_r$ or $E_r = 0$. We then check whether the graph is fragmented or not (see README on \href{https://github.com/quentinrsl/capsidgraph}{Gihub}), and compute the fragmentation threshold of such a capsid using the bisection method described below \ref{m:general}.

Similarly, for node removal, we first compute the energy of each node by adding up the bond energies of each edge connected to it, and then compute its reciprocal to obtain its probability weight. We again choose an amount of energy to randomly remove ($E_r$) and randomly select a node for removal. For each edge connected to the chosen node, we subtract its bond energy from the energy of its neighbouring nodes. If a node is now isolated, i.e. its energy is zero, it is removed from the graph and its energy subtracted from $E_r$. We stop this process once the energy of each remaining node is greater than $E_r$, and then check if the graph is fragmented. Fragmentation under edge removal is then again determined with the same algorithm as in \ref{m:general}.

\subsection{The bisection method} \label{m:general}
\begin{figure}[H]
    \includegraphics[width=0.6\linewidth]{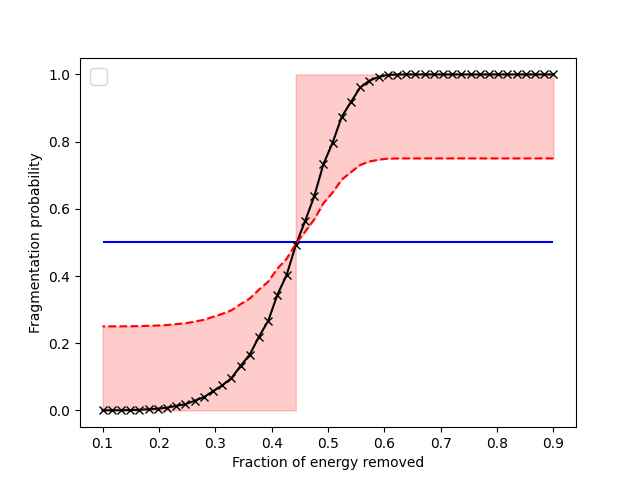}
    \centering
    \caption{A step in the bisection method stops when the probability of $\frac{S_N}{N}$ being in the red area is above $1-\epsilon$.}
    \label{fig:bisectioninequality}
\end{figure}

To determine the fragmentation threshold, we use a bisection method. For each step of the algorithm, we determine whether the probability of fragmentation $p_f$ is above or below $0.5$ with a certain accuracy, i.e., with a high enough probability. For this, let $N$ be the number of simulations, $\epsilon$ the upper bound for the probability of having a wrong value for the next step (i.e. for getting a value above 0.5 were the actual one is below or vice versa). Let $F(f_r)$ be a random variable which returns $1$ if the graph is fragmented after removing a node/edge with probability $f_r$, or $0$ otherwise. This variable has a Bernoulli distribution $F(f_r) \sim \mathrm{B}(p_f)$. Let $(F_i)_{i\in[1,N]}$ be N independent variables such that $\forall i\in[1,N], F_i \sim \mathrm{B}(p_f)$, then $S_N = \sum_{i=1}^{N}{F_i} \sim \mathrm{B}(N,p_f)$. $S_N$ is a new random variable that represents the number of simulations that resulted in a fragmented capsid after N tries. We know that $\mathrm{E}(\frac{S_N}{N})=p_f$. Chebyshev's inequality then yields $\forall a>0$:
\begin{equation}
\label{eq:chebitchev}
\begin{split}
    \textbf{P}(|\frac{S_N}{N} - p_f| > a) &\leq \frac{\textbf{V}(\frac{S_N}{N})}{a^2}\\
    &= \frac{Np_f(1-p_f)}{N^2 a^2} \leq \frac{1}{4Na^2}
\end{split}
\end{equation}
If $|\frac{S_N}{N} - p_f| < |S_N - 0.5|$, then $\frac{S_N}{N}$ lies in the red area in figure  \ref{fig:bisectioninequality}, i.e. closer to the black than the blue curve, implying that $S_N$ is in the correct range for the next step of the bisection method, and we therefore stop the simulation at this point.

This gives us
\begin{eqnarray}
    \textbf{P}(\mathrm{error}) \leq \textbf{P}(|\frac{S_N}{N} - p_f| > |S_n - 0.5|)
\end{eqnarray}
By applying \ref{eq:chebitchev} with $a=|S_n - 0.5|$ we get
\begin{eqnarray}
    \textbf{P}(|\frac{S_N}{N} - p_f| > |S_n - 0.5|) \leq \frac{1}{4N|S_N - 0.5|^2}\,,
\end{eqnarray}
hence
\begin{eqnarray}
    4N|\frac{S_N}{N}-0.5|^2>\frac{1}{\epsilon} \implies \textbf{P}(\mathrm{error}) < \epsilon
\end{eqnarray}

This inequality defines the stop condition for each step of the bisection method. As long as $\lim_{N \to \infty}\frac{S_N}{N} \neq 0.5$ the algorithm will stop. However, the number of iterations this will take is potentially unbounded. Therefore, a maximal number of iterations is set at which the bisection process terminates. In none of our simulations that value was ever reached.

\subsection{Definition of the largest hole size}
For algorithmic purposes, we need a formal definition of the largest hole size.
\begin{definition}
\label{def:hole_size}
Let $G = (V,E)$ be a connected graph, and $G' = (V',E')$ a subgraph of $G$ where $G \neq G'$ and $V' \neq \emptyset$. 
Consider the set of connected components of maximal size (i.e., with the largest number of nodes) $\{C_0,...,C_{p-1}\}$. 
Let $i \in \{0,...,p-1\}$, $C_i=(V_i,E_i)$ and $\bar{C_i}=(\bar{V_i}, \bar{E_i})$ where
$\bar{V_i} = V \setminus V_i$ and $\bar{E_i} = \{\{u,v\} :\{u,v\}\in E, u \in \bar{V_i}, v \in \bar{V_i}\}$. Further, let $H_i$ be the size of the largest connected component of $\bar{C_i}$.\\
Then the hole size of $G'$ is defined as
$$H_G(G') = \max_{0 \leq j \leq p-1 }{H_j}\,.$$
By convention, we set $H_G(G) = 0$ and $H_G(\emptyset)$ = |V|.
\end{definition}

\begin{figure}[H]
    \centering
    \includegraphics[width=0.6\linewidth]{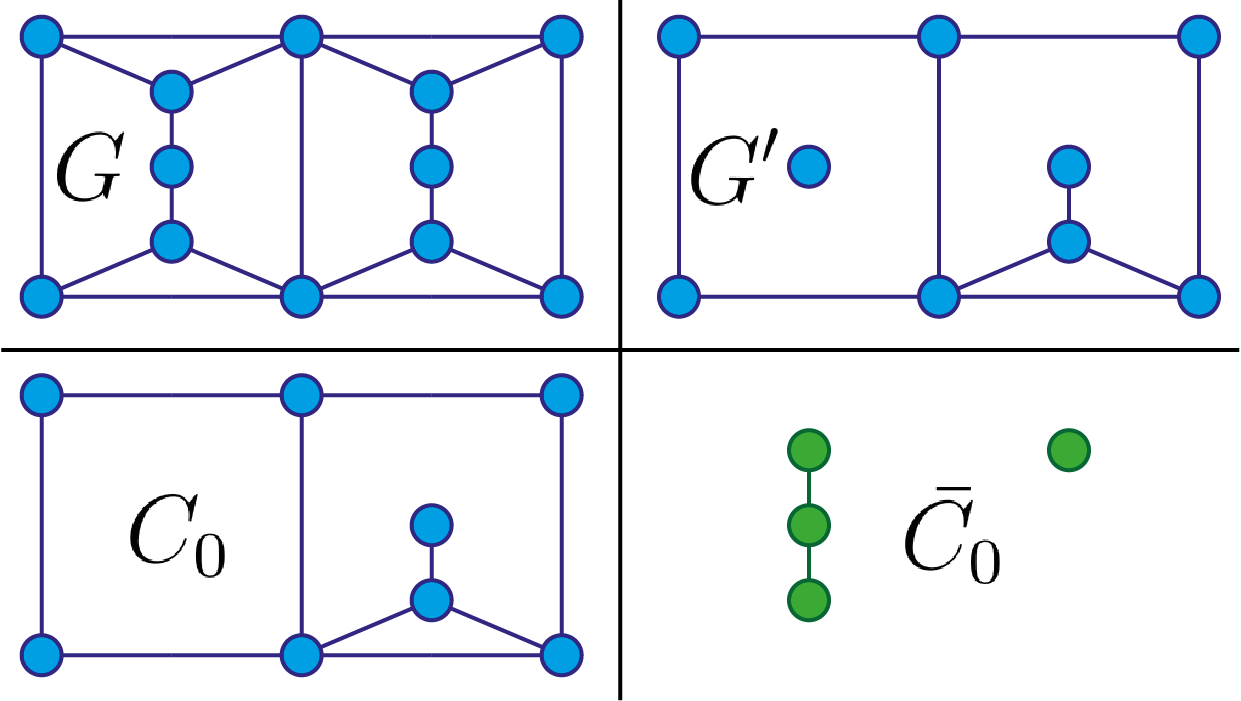}
    \caption{An example of a graph $G'$ with two connected components. The set of connected components contains one subgraph $C_0$. We have $H_0=3$, hence $H_G(G')=3$.}
    \label{fig:example_hole_size}
\end{figure}

Some instructive examples illustrate the rationale underpinning this definition. In order to describe the size of the largest hole in the bulk ("main component") of the capsid, one approach would be to compute the size of the largest connected component of $G \setminus G'$. However, note that this definition would find the graph of Fig.\ref{fig:example_hole_size} as having a hole size of 1, because the isolated node of $G'$ is still considered part of the graph, even though it is no longer part of the "main component" that corresponds to the bulk of the capsid. 

\begin{figure}[H]
    \centering
    \includegraphics[width=0.6\columnwidth]{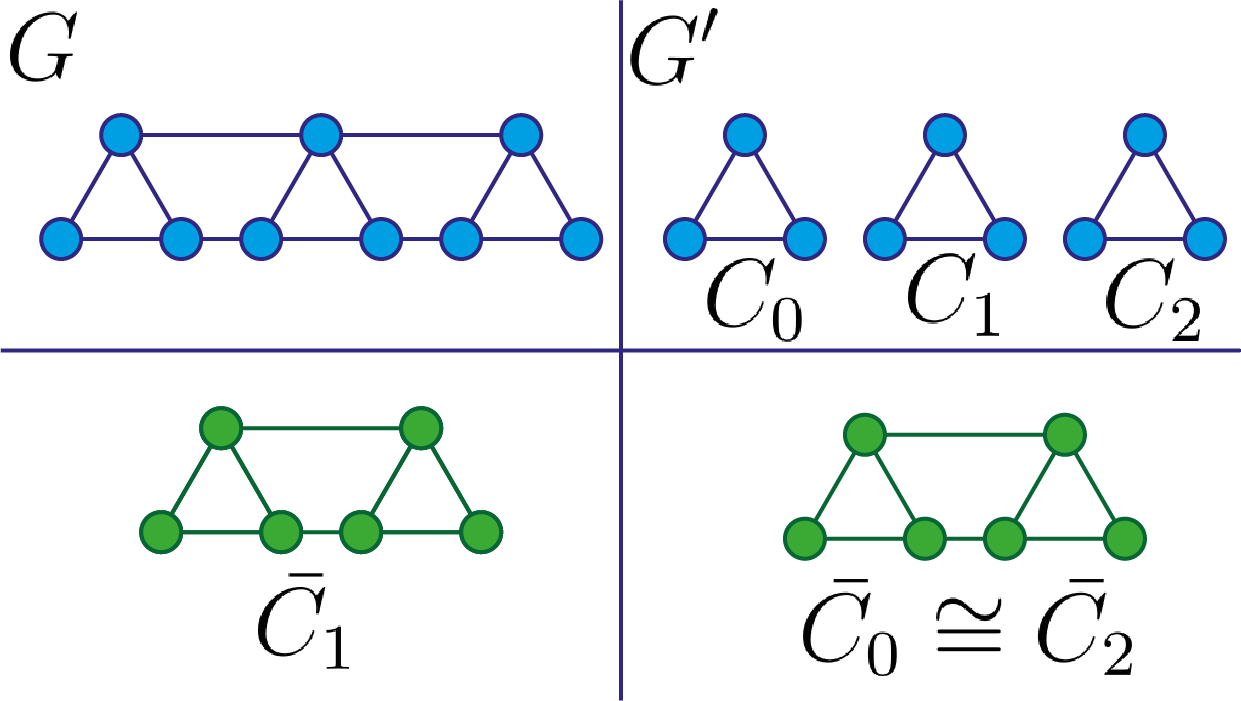}
    \caption{An example of a graph $G'$ with three connected components with the same number of vertices$\{C_0,C_1,C_2\}$. We have $H_0=H_2=6$ and $H_1=3$. Thus, $H_G(G')=6$. }
    \label{fig:example_hole_size_2}
\end{figure}

For this reason we only consider the largest connected component. We denote by $\bar{C_0}$ the graph made of the "missing" pieces from $C_0$, i.e. the graph corresponding to the "holes" in $C_0$. In case there are 
multiple largest connected components as in Fig. \ref{fig:example_hole_size_2}, the algorithm has to decide which to pick. 
Intuitively, this is equivalent to choosing which is the main component. This can happen in practice for instance if a capsid graph breaks into three equal-sized pieces with a "middle ring" connecting two "disks". The question we need to ask is whether we consider such a graph as having two holes 1/3 of the capsid size, or one hole 2/3 of the capsid size. By using $H_G(G') = \max_{0 \leq j \leq p-1 }{H_j}$ in Def. \ref{def:hole_size}, we opt for the latter case. 

\begin{figure}[H]
    \centering
    \includegraphics[width=0.6\columnwidth]{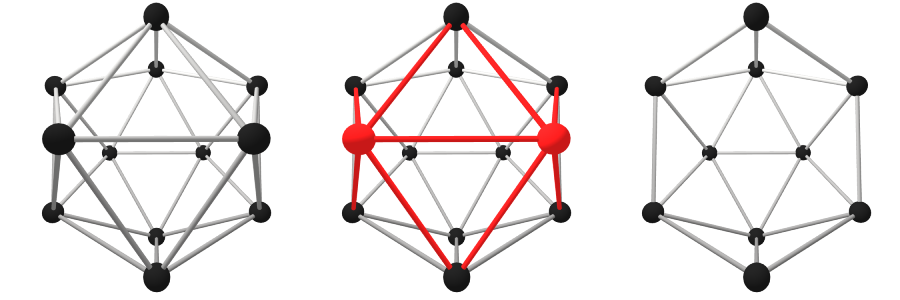}
    \caption{Removal of two nodes (red) results in a hole of size two in the graph shown.}
    \label{fig:holesize_definition_example}
\end{figure}

However, we note that these cases are rare. Typically, we can easily determine the size of the largest "hole" present in the capsid by considering the largest connected component of the fragmented capsid as the "main part" or "bulk" of the capsid. Any group of neighbouring missing subunits would then be a "hole", and the largest group would correspond to the largest hole, as illustrated by an example in Fig. \ref{fig:holesize_definition_example}.
The probability distribution of largest hole size for a given fragmentation energy shows the tendency of the graph to either break apart completely, or only exhibit small missing fragments.  As expected, when removing small values of energy, the hole sizes tend to be consistently small. On the other hand, when removing most of the capsid energy, the largest hole tends to consist of most of the capsid. However, the transition between these two regimes is not linear and happens abruptly for a specific energy value $E_H$. This value can be formally defined as the removal energy for which the probability of the largest hole being larger than half of the capsid is 0.5. This values can be interpreted as the energy needed to break the structure of the capsid and is a measure of the graph's resilience to fragmentation.

\subsection{The entropy of the hole distribution in disassembly intermediates}

The randomness of each distribution can be quantitatively estimated using its entropy. For a capsid of size $n$, with a hole size ranging from 0 to $n$, this entropy ranges from 0 for the distribution of a deterministic random variable (i.e., the hole size is always the same for this distribution) to $\log_{2}(n+1)$ for a uniform distribution over all hole sizes. This entropy value $H$ is observed to be maximal for $E_r=E_H$. For this value to be comparable between graphs of different sizes, we need to normalize it by $\log_{2}(n+1)$.

\subsection{Simulation parameters}

When computing fragmentation and hole formation probabilities for given values of energy removed ($E_r$), the only free parameter is the number of Monte Carlo steps. Estimation of fragmentation and hole formation thresholds is done using a bisection method, which is characterized by its number of steps, the probability of error in each step, and the maximal number of simulations per step. These parameters are given in Tables \ref{Tab1} \& \ref{Tab2}. 
\begin{center}
    \begin{tabularx}{0.55\linewidth}{|X|X|}
        \hline
        Figure & Iterations \\
        \hline
        Fig.\ref{fig:frag_prob_polyoma_edges} \& \ref{fig:frag_prob_polyoma_nodes} & 1,000,000 / point \\
        \hline
        Fig.\ref{fig:comparaison_polyoma} \& \ref{fig:comparaison_other} & 100,000 / point \\
        \hline
        Fig.\ref{fig:polyoma_sizeholedist} & 1,000,000 / distribution \\
        \hline
    \end{tabularx}
    \label{Tab1}
    \captionof{table}{Computational setting for Monte Carlo simluations with a fixed number of iteration steps.}
\end{center}
    
\begin{center}
    \begin{tabularx}{0.75\linewidth}{|X|X|X|X|}
        \hline
        Figure & Bisection steps & Error probability & Maximum iteration per step \\
        \hline
        fig.\ref{fig:landscape_polyoma_edges} \& \ref{fig:landscape_polyoma_nodes} & 8 & 0.05 & 10,000,000 \\
        \hline
        fig.\ref{fig:CKResults} & 9 & 0.05 & 5,000,000 \\
        \hline
        fig.\ref{fig:aals} & 9 & 0.01 & 1,000,000,000 \\
        \hline
    \end{tabularx}
    \label{Tab2}
    \captionof{table}{Computational setting for values estimated through the bisection method.}
\end{center}

\section{Conclusion}

This comparative analysis of viral and \textit{de novo} designed protein cage architectures of similar size reveals different propensities for fragmentation for distinct capsid architectures. A comparison of different viral cages -- quasiequivalent CK geometries and non-quasiequivalent papillomavirus cages -- shows comparable properties, albeit with the non-quasiequivalent capsid being more prone to fragmentation. In both types of viral capsid architecture, disassembly pathways are more likely to occur via hole formation than via capsid fragmentation. 

By contrast, the 72-pentamer AaLs cage is more likely to disassemble via fragmentation. This trend is shared also by the smaller AaLS cages, suggesting that it is a common property of these \textit{de novo} designed cages. This is likely due to the fact that in contrast to viral capsids, some protein subunits in their capsomers do not interact with other capsomers in the cage, leading to the formation of larger holes in the cage surface. Interestingly, a similar behaviour is seen also in viruses formed from 72 capsomers (12 pentamers and 60 hexamers) that are organised according to a rhomb tiling, as for example in bacteriophage Hong-Kong 97 (HK97). This might explain why these viruses have evolved additional capsid features, such as the chain-mail organisation in HK97 \cite{chain_mail}, to stabilise their capsids.

In summary, different capsid architectures follow principally different disassembly mechanisms, with a preference for either hole formation or fragmentation. Our analysis shows evidence of both in naturally occuring viruses depending on their geometric design principles. These results provides a guide for protein nanoparticle design targeted at specific applications, contributing to the rational design of specific desired cargo release mechanisms. 

\section*{Acknowledgements}

RT thanks the Wellcome Trust for financial support through the Joint Investigator Award (110145 \& 110146), the EPSRC for an Established Career Fellowship (EP/R023204/1) which also provided funding for QR and the Royal Society for a Royal Society Wolfson Fellowship (RSWF/R1/180009), which provided funding for QR and SB.



\end{document}